\documentclass[fleqn,twoside]{article}
\usepackage{espcrc2}

\usepackage[dvips]{graphicx}

\title{
 \vspace{-1.8cm}
 \hfill \rm \null \hfill
 \hbox{\normalsize ADP-02-85/T524} \\
 \vspace{+1.3cm}
FLIC-Overlap Fermions and Topology}

\author{W. Kamleh\address[CSSM]{Special Research Centre for the Subatomic Structure of Matter (CSSM) and Department of Physics and Mathematical Physics, University of Adelaide 5005, Australia.},
D.J. Kusterer\address[Tub]{Institut f\"ur Theoretische Physik,
Universit\"at T\"ubingen, D-72076, Germany},
D.B. Leinweber\addressmark[CSSM]\thanks{Presented by D.B. Leinweber at Lattice '02.},
A.G. Williams\addressmark[CSSM]
}

\begin{document}

\thispagestyle{empty}

\begin{abstract}
APE smearing the links in the irrelevant operators of clover fermions
(Fat-Link Irrelevant Clover (FLIC) fermions) provides significant
improvement in the condition number of the Hermitian-Dirac operator
and gives rise to a factor of two savings in computing the overlap
operator.  This report investigates the effects of using a
highly-improved definition of the lattice field-strength tensor
$F_{\mu \nu}$ in the fermion action, made possible through the use of
APE-smeared fat links in the construction of the irrelevant operators.
Spurious double-zero crossings in the spectral flow of the
Hermitian-Wilson Dirac operator associated with lattice artifacts at
the scale of the lattice spacing are removed with FLIC fermions
composed with an ${\cal O}(a^4)$-improved lattice field strength
tensor.  Hence, FLIC-Overlap fermions provide an additional benefit to
the overlap formalism: a correct realization of topology in the
fermion sector on the lattice.
\end{abstract}

\maketitle

\section{SPECTRAL FLOW}

Overlap fermions \cite{overlap4} are a realisation of chiral symmetry
on the lattice.  Given some reasonable Hermitian-Dirac operator $H$, we
can deform $H$ into a chiral action through the overlap formalism,
\begin{equation}
D_o = \frac{1}{2}\big( 1+\gamma_5 \, \epsilon(H) \big), \quad
\epsilon(H)=\frac{H}{\sqrt{H^2}} \, ,
\end{equation}
and $H$ will be referred to as the overlap kernel.

\begin{figure}[!tb]
\begin{center}
\includegraphics[width=0.45\textwidth, angle=0 ]{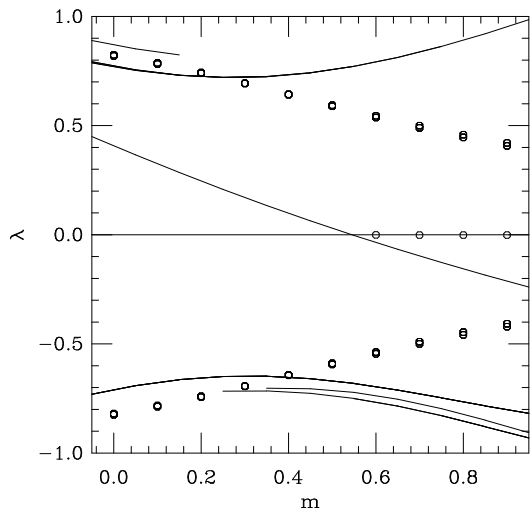}
\end{center}
\vspace{-48pt}
\caption{Low-lying eigenmode flow of the Hermitian-Wilson kernel
  (curves) and the flow of the associated overlap operator (circles)
  from Ref.~\protect\cite{edwards-practical}.
  \vspace{-24pt}}
\label{fig:connection}
\end{figure}

Examinations of the spectral flow of the overlap kernel can shed
considerable light on the nature of the associated overlap operator.
Consider for example the spectral flow of Figure \ref{fig:connection}
for overlap fermions based on the standard Wilson kernel, reproduced
from Ref.~\cite{edwards-practical}.  As the regulator-mass parameter
of the Wilson kernel is varied over the doubler-free region ($0 < m <
2$ at tree level) zero crossings occur, signaling the point at which
the associated overlap operator becomes sensitive to the topology
giving rise to the zero mode.  The sign of the slope of the kernel
flow at the zero crossing indicates the sign of the topological charge
giving rise to the zero mode.  As such the flow should only cross zero
once.  

The position of the zero crossing reflects the size of the underlying
topological object \cite{Kusterer:2001vk}.  Smaller objects give rise
to zero crossings at larger regulator mass $m$.

\section{SPURIOUS DOUBLE CROSSINGS}

\begin{figure*}[!tb]
\includegraphics[height=0.48\textwidth,angle=90]{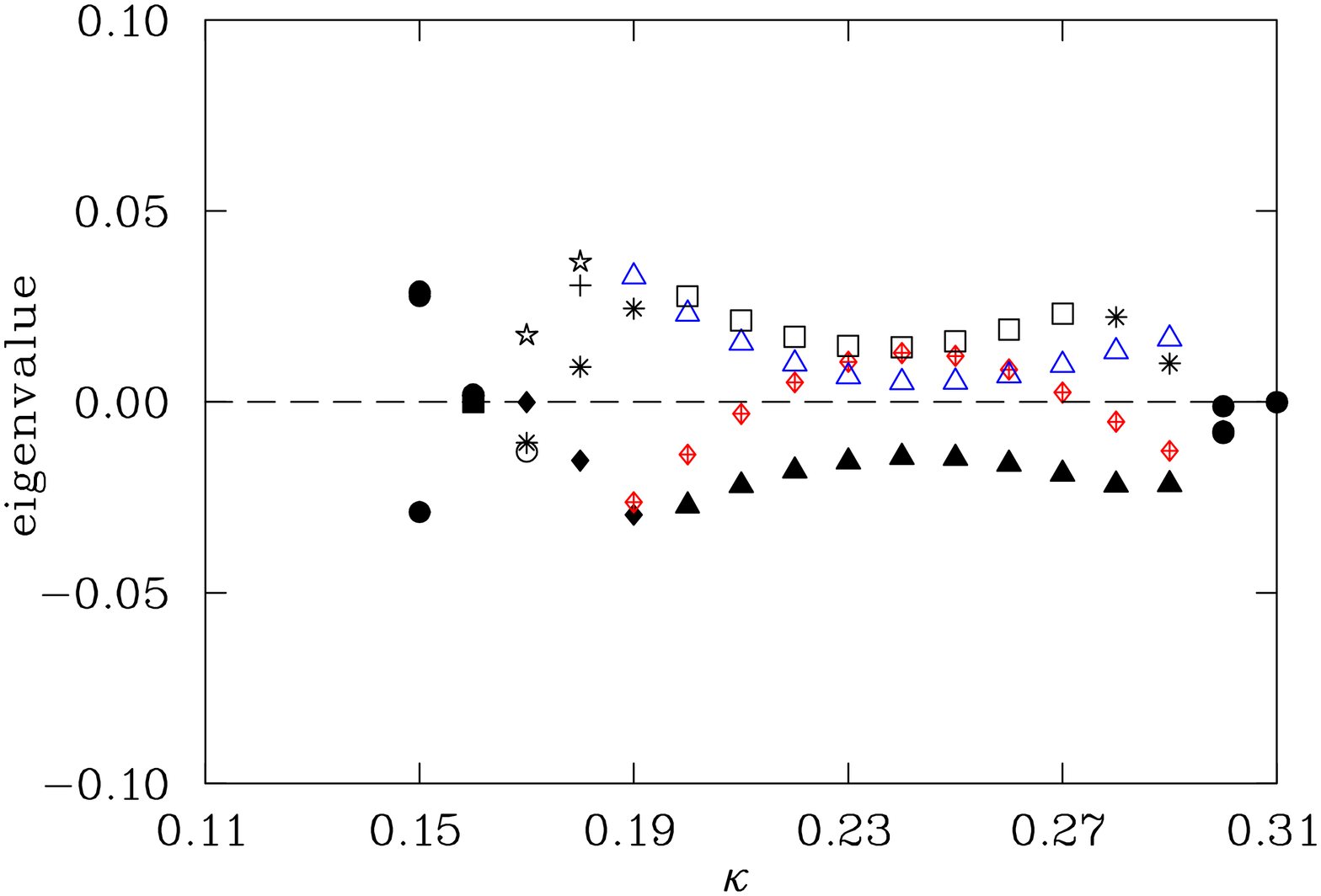}
\includegraphics[height=0.48\textwidth,angle=90]{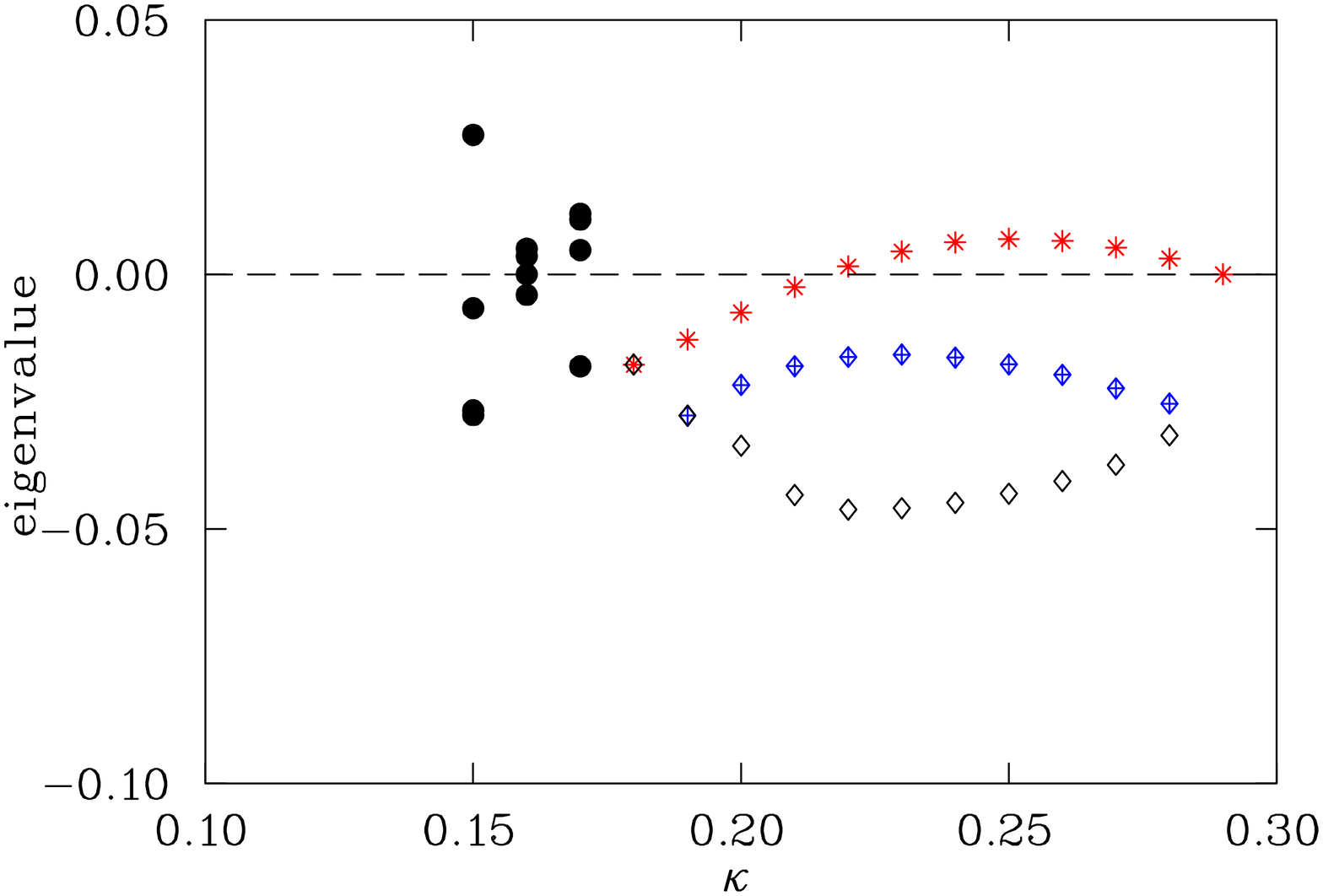}
\vspace{-24pt}
\caption{Low-lying eigenmode flow of the standard Hermitian-Wilson
  kernel for two selected configurations displaying double crossings
  of a single mode.
\vspace{12pt}
\label{fig:Wilson}}
\end{figure*}

In our examinations of the spectral flow of the standard Wilson kernel
\cite{Kusterer:2001vk} and the correlation of the eigenmodes with
their underlying topological gauge field structures, we encountered a
few configurations in which the spectral flow of a single mode would
cross zero twice in the doubler free region.  Figure~\ref{fig:Wilson}
illustrates the low-lying eigenmode spectrum for two configurations
which display a double crossing in one of the modes.  The relation
between the hopping parameter $\kappa$ and the regulator mass of
Fig.~\ref{fig:connection} is $\kappa = { 1 / ( -2\, m + 8 \, r ) }$
such that the tree-level doubler-free region is $1/8 < \kappa < 1/4$
for $r = 1$.  The $SU(3)$ gauge-field configurations are generated on
a $16^3 \times 32$ lattice with a mean-field improved plaquette plus
rectangle action at $\beta = 4.60$ providing a lattice spacing of
0.122(2) fm.  Modes are tracked in the spectral flow by examining
their position in space-time.  Modes belonging to the same underlying
topological structure are plotted with a common symbol in the spectral
flow plots.  Our experience is that about one in five configurations
displays such a double crossing.

\begin{figure*}[!tb]
\includegraphics[height=0.48\textwidth,angle=90]{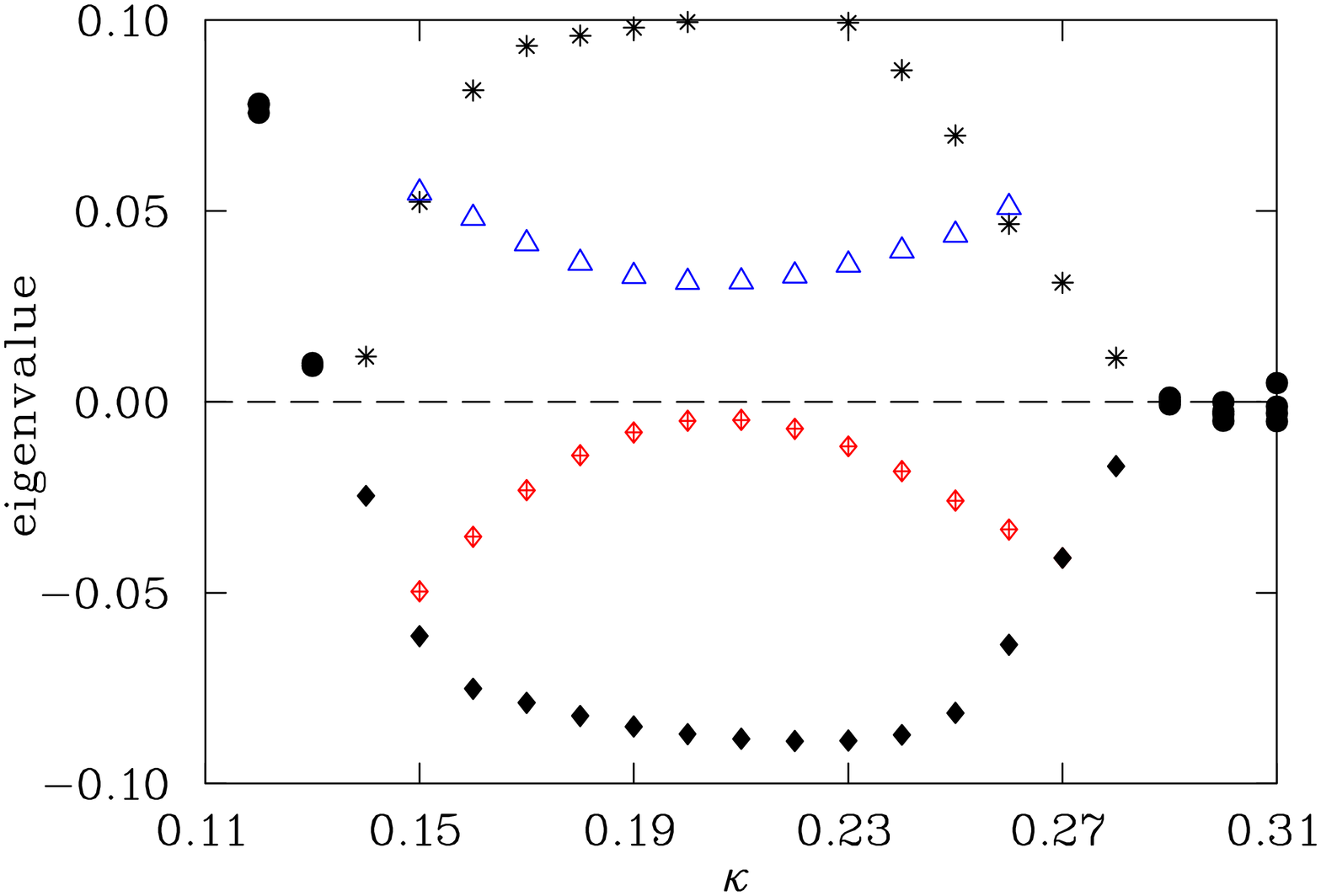}
\includegraphics[height=0.48\textwidth,angle=90]{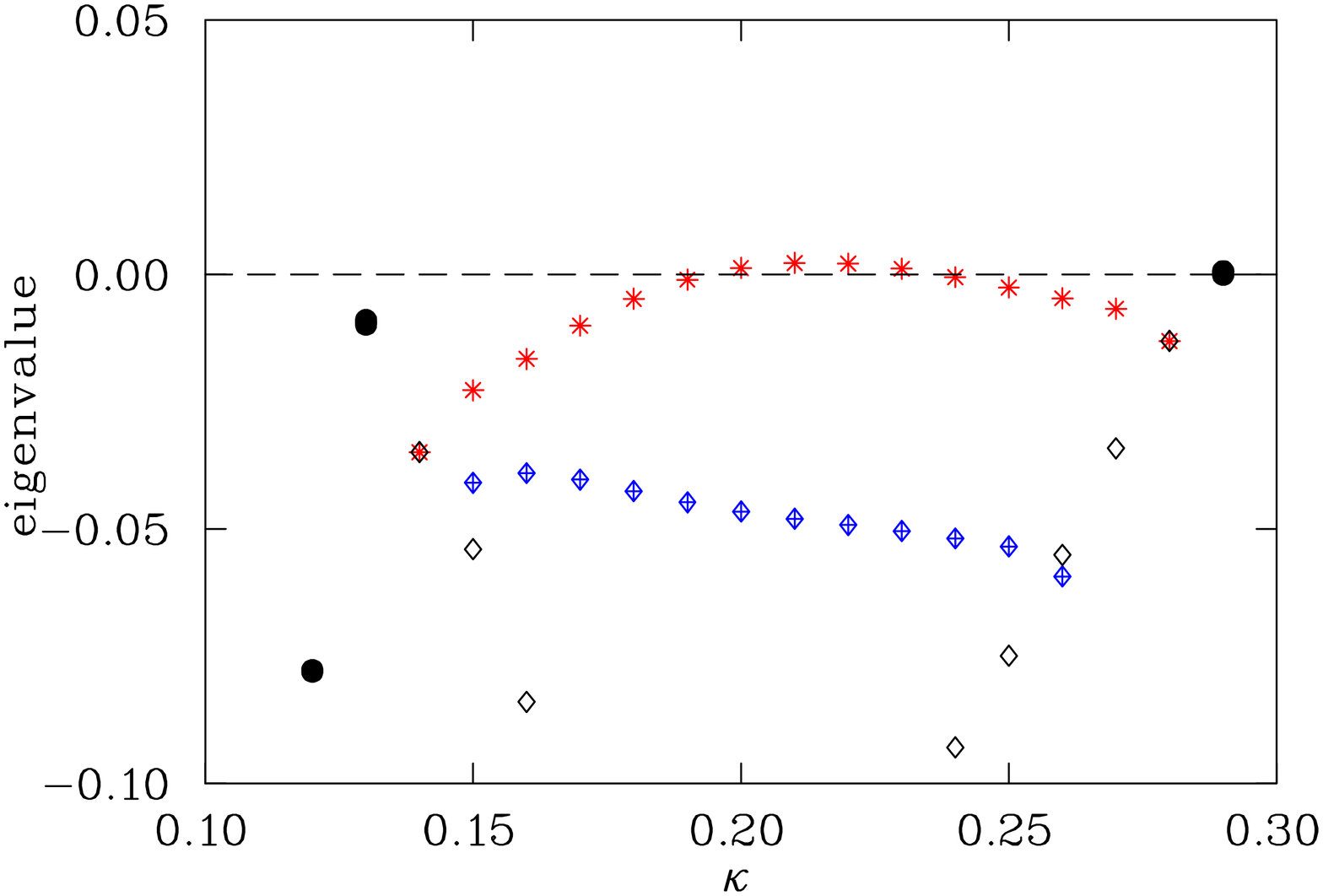}
\vspace{-24pt}
\caption{Low-lying eigenmode flow of the Hermitian-FLIC kernel with
  $F_{\mu \nu}$ estimated via the standard clover-plaquette links
  paths.  Selected configurations are as in Fig.~\protect\ref{fig:Wilson}.
\vspace{12pt}
\label{fig:1LIC_FLIC}}
\end{figure*}

Figure~\ref{fig:1LIC_FLIC} displays the spectral flow for the same
configurations but this time evaluated with the Fat-Link Irrelevant
Clover (FLIC)-fermion kernel \cite{zanotti-hadron,kamleh-overlap}
constructed with four-sweep APE-smeared fat links with a smearing
fraction \cite{Bonnet:2000dc} of $\alpha = 0.7$.  FLIC fermions
provide an effective alternative to the Wilson kernel as they have an
improved condition number which gives rise to a factor of two reduction
in compute time.  Here the double crossing is removed for one
configuration but survives for the other.

\begin{figure*}[!tb]
\includegraphics[height=0.48\textwidth,angle=90]{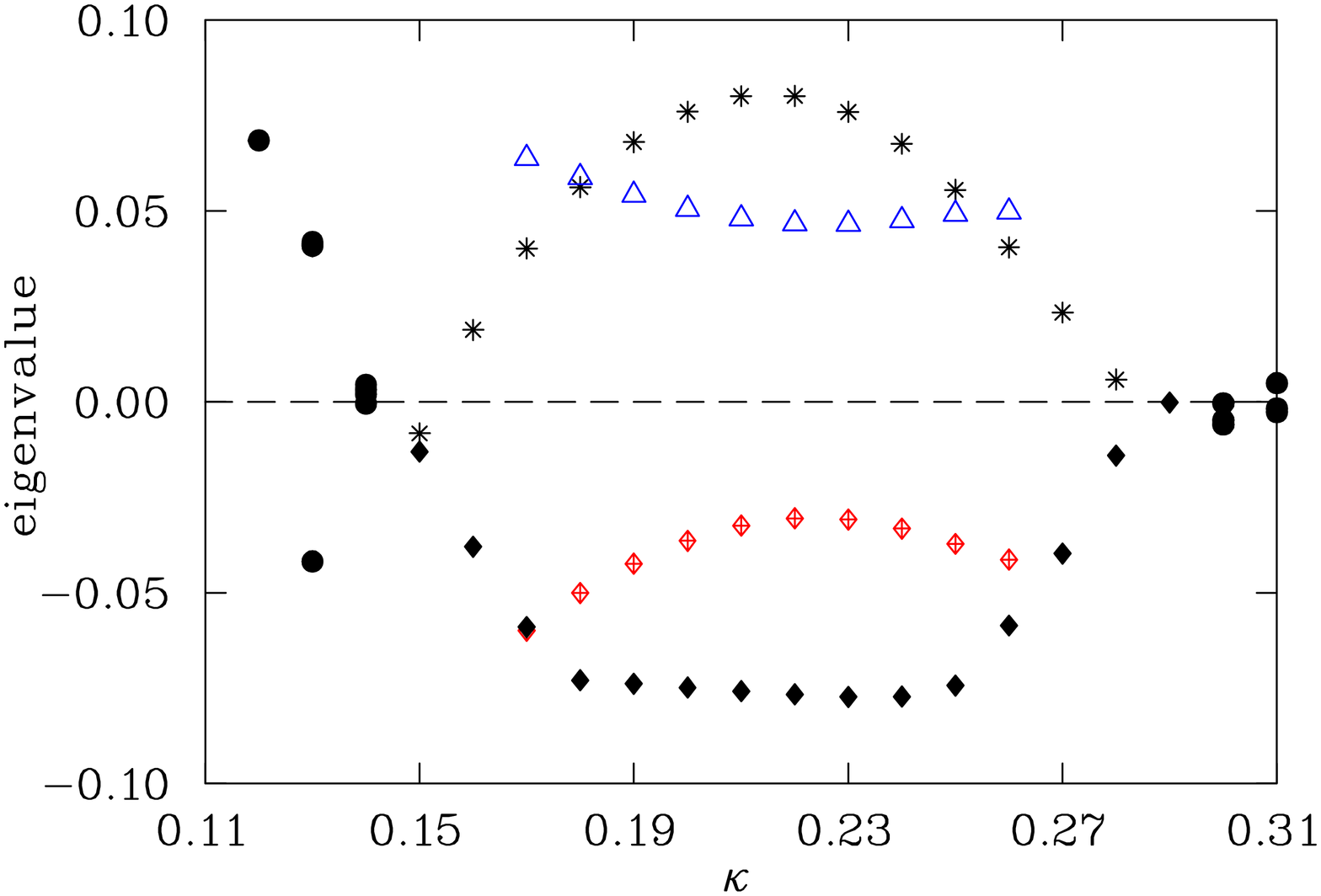}
\includegraphics[height=0.48\textwidth,angle=90]{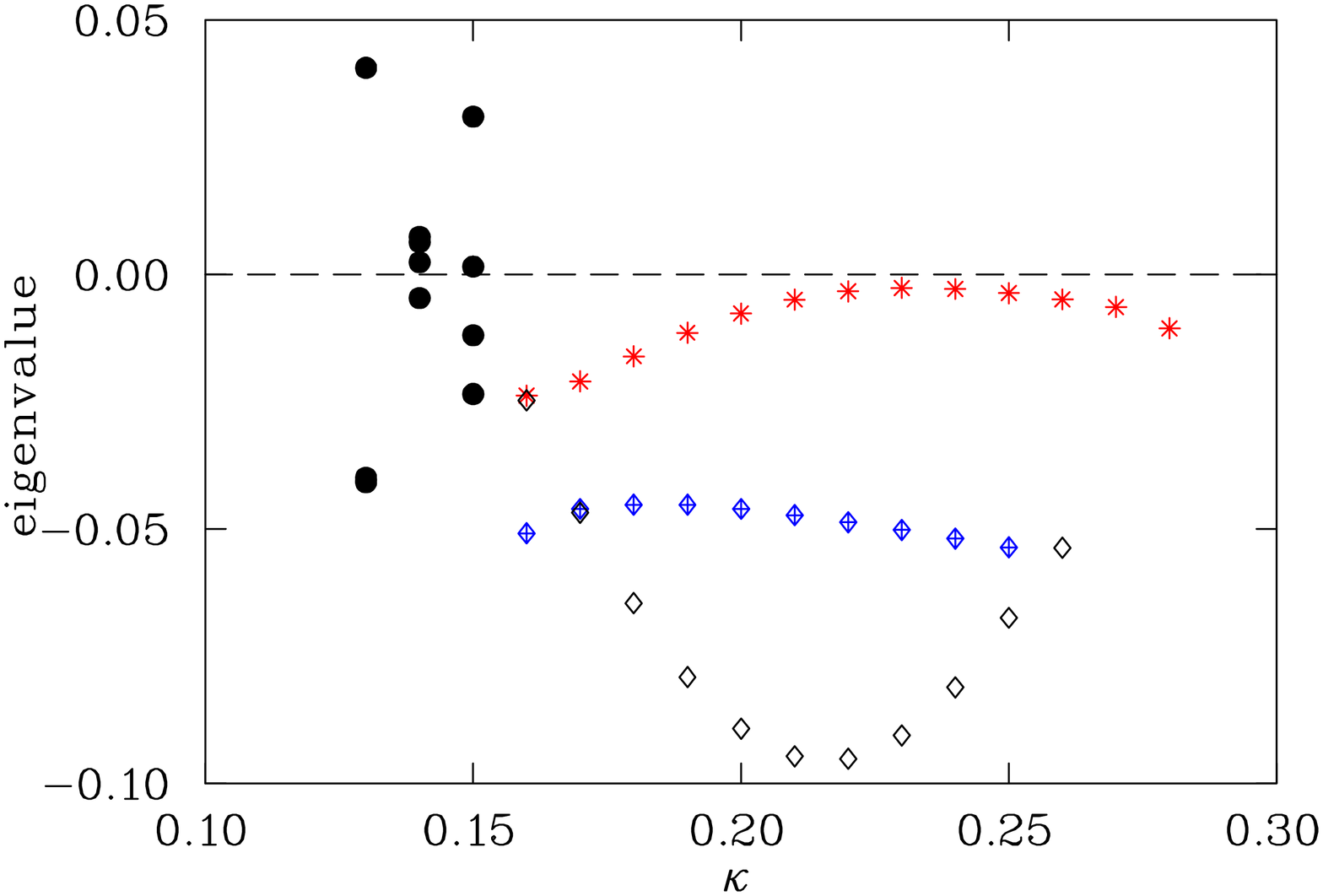}
\vspace{-24pt}
\caption{Low-lying eigenmode flow of the Hermitian-FLIC kernel with
  $F_{\mu \nu}$ determined via the ${\cal O}(a^4)$-improved three-loop
  definition \protect\cite{Bilson-Thompson:2002jk}.  Selected
  configurations are as in
  Figs.~\protect\ref{fig:Wilson}~and~\protect\ref{fig:1LIC_FLIC}.
\label{fig:3LIC_FLIC}}
\end{figure*}

The plaquette-based clover estimate of the lattice field-strength
tensor $F_{\mu \nu}$ is known to be rather poor.  The associated
topological charge typically differs from integer values by 10\%, even
on very smooth configurations \cite{Bonnet:2000dc}.  Since the
irrelevant operators of FLIC fermions are constructed from APE-smeared
links, one may use highly-improved ${\cal O}(a^4)$-improved
definitions of the lattice field strength tensor
\cite{Bilson-Thompson:2002jk}; ${\cal O}(g^2\, a^2)$ errors are
removed in the APE-smearing process.  Figure~\ref{fig:3LIC_FLIC}
illustrates the spectral flow obtained from FLIC fermions
incorporating a three-loop ${\cal O}(a^4)$-improved definition of the
lattice field strength tensor \cite{Bilson-Thompson:2002jk}.  The
spurious double crossings are eliminated.  It is perhaps interesting
to note that the double-crossings can be eliminated with the Wilson
kernel provided the irrelevant Wilson term is constructed with fat
links.  However, a factor of two speedup is not realized for the
fat-Wilson kernel.

\section{CONCLUSION}

We have been successful in creating spectral flows with a spurious
double crossing \cite{kustererSoon} by cooling t'Hooft-ansatz
instantons on the lattice until their topological charge is reduced to
0.4.  This suggests to us that the observed double crossings seen with
the standard Wilson kernel are simply lattice artifacts associated
with dislocations.  The fact that the zero crossings always occur at
very large values of the regulator parameter further confirms that
these double crossings are spurious lattice artifacts due to physics
at the scale of the lattice spacing and should be removed.  The
three-loop improved FLIC fermion kernel achieves this goal.

%


\begin{thebibliography}{10}

\bibitem{overlap4}
R. Narayanan and H. Neuberger,
\newblock Nucl. Phys. B443 (1995) 305, hep-th/9411108.

\bibitem{edwards-practical}
R.G. Edwards, U.M. Heller and R. Narayanan,
\newblock Nucl. Phys. { B540} (1999) 457, hep-lat/9807017.

\bibitem{Kusterer:2001vk}
D.~J.~Kusterer, J.~Hedditch, W.~Kamleh, D.~B.~Leinweber and A.~G.~Williams,
Nucl.\ Phys.\ B {\bf 628} (2002) 253
[arXiv:hep-lat/0111029].

\bibitem{zanotti-hadron}
CSSM Lattice, J.M. Zanotti et~al.,
\newblock Phys. Rev. D65 (2002) 074507, hep-lat/0110216.

\bibitem{kamleh-overlap}
W. Kamleh et~al.,
\newblock Phys. Rev. D66 (2002) 014501, hep-lat/0112041.

\bibitem{Bonnet:2000dc}
F.~D.~Bonnet, P.~Fitzhenry, D.~B.~Leinweber, M.~R.~Stanford and A.~G.~Williams,
Phys.\ Rev.\ D {\bf 62} (2000) 094509
[arXiv:hep-lat/0001018].

\bibitem{Bilson-Thompson:2002jk}
S.~O.~Bilson-Thompson, D.~B.~Leinweber and A.~G.~Williams,
arXiv:hep-lat/0203008.

\bibitem{kustererSoon}
D. Kusterer, {\it et al.}, in preparation.

\end{thebibliography}
\end{document}